\documentclass[onecolumn,10pt]{article}

\usepackage{amssymb}
\usepackage[T1]{fontenc}
\usepackage{times}
\newcommand{\alg}[1]{\mbox{$\mathfrak{#1}$}}

\newcommand{\hil}[1]{\mbox{$\mathcal{#1}$}}

\begin{document}

\title{\textbf{Quantum Mechanics is About Quantum Information}}
 \vspace{1in}

\author{\textbf{Jeffrey Bub}\\
Department of Philosophy, University of Maryland, \\
College Park, MD 20742 \\
(E-mail: jbub@carnap.umd.edu)}
\vspace{.5in}

\maketitle

\begin{abstract}
I argue that quantum mechanics is fundamentally a theory about the 
representation and manipulation of information, not a 
theory about the mechanics of nonclassical waves or particles. 
The notion of quantum information is to be understood as a new 
physical primitive---just as, following Einstein's special theory of 
relativity, a field is no longer regarded as the physical 
manifestation of vibrations in a mechanical medium, but recognized as 
a new physical primitive in its own right.
\end{abstract}  

\bigskip

\section{Introduction}

In several places \cite{Cushing1,Cushing2,Cushing3}, Cushing 
speculates about the possibility of an 
alternative history, in which Bohm's theory \cite{BohmHiley, 
Goldstein} is developed as the 
standard version of quantum mechanics, and suggests that in that 
case the Copenhagen interpretation, if it had been proposed as an 
alternative to a fully developed Bohmian theory, 
would have been summarily rejected. I quote from \cite[pp. 
352--353]{Cushing2}:
\begin{quote}
\ldots we can fashion a highly reconstructed but entirely plausible 
bit of 
partially `counterfactual' history as follows (all around 
1925--1927). 
Heisenberg's matrix mechanics 
and Schr\"{o}dinger's wave mechanics are formulated and shown to be 
mathematically equivalent. Study of a classical particle subject to 
Brownian motion \ldots leads to 
a classical understanding of the already discovered 
Schr\"{o}dinger equation. A stochastic mechanics underpins this 
interpretation 
with a visualizable model of microphenomena and, so, a realistic 
ontology
remains viable.
Since stochastic mechanics is quite difficult to handle 
mathematically, study naturally turns to the mathematically 
equivalent 
linear
Schr\"{o}dinger equation. Hence, the Dirac transformation theory and 
an 
operator formalism are available as a 
convenience for further development of the mathematics to 
provide algorithms for calculation.

\ldots

A Bell-type theorem is proven and taken as convincing evidence that 
nonlocality is present in quantum phenomena. A no-signalling theorem 
for quantum mechanical correlations is established and this puts to 
rest Einstein's objections to the nonseparability of quantum 
mechanics. \ldots This could reasonably have been enough to overcome 
his objections to the nonlocal nature of a de Broglie--Bohm 
interpretation of the formalism of quantum mechanics. Because the 
stochastic theory is both nonlocal and indeterministic, whereas the 
de Broglie--Bohm model is nonlocal only and still susceptible to a 
realist interpretation, Einstein might have made the transition to 
the 
latter type of theory. 

That is, these developments, that could, conceptually and logically, 
have taken place around 1927, could have overcome the resistance of 
Einstein and of Schr\"{o}dinger to supporting a de Broglie--Bohm 
program. \ldots Bohm's interpretation would certainly have been 
possible in 1927. These models and theories could be generalized to 
include relativity and spin. The program is off and running. Finally, 
this causal interpretation can be extended to quantum fields. 

So, if, say, in 1927, the fate of the causal interpretation had taken 
a very different turn and been accepted over the Copenhagen one, it 
would have had the resources to cope with the generalizations 
essential for a broad-based empirical adequacy. We could today have 
arrived at a very different world view of microphenomena. If someone 
were then to present the merely empirically equally as adequate 
Copenhagen version, with all of its own additional counterintuitive 
and mind-boggling aspects, who would listen? \ldots However, 
Copenhagen got to the top of the hill first and, to most practicing 
scientists, there seems to be no point in dislodging it.
\end{quote}

Cushing's broader and very interesting thesis was that the successful 
theories that 
philosophers of science analyze as case-studies
 are themselves contingent on historical 
factors---in particular, the success of the
Copenhagen interpretation of quantum mechanics is a 
matter of historical contingency. 
I want to play Cushing's counterfactual game for the case of 
special relativity and compare this with the quantum mechanical case to argue
 for a very different thesis: the interpretation of 
quantum mechanics as a theory about the 
representation and manipulation of information in our world, not a 
theory about the mechanics of nonclassical waves or particles. 

The following discussion is divided into three sections. In `Principle 
vs Constructive Theories,' I discuss Einstein's distinction between 
these two classes of theories, and the significance of his characterization of 
special relativity as a principle theory. I conclude the section by 
arguing that,
just as the rejection of Lorentz's theory in favour of special 
relativity (formulated in terms of Einstein's two principles)
 involved taking the notion of a field as a new physical 
primitive, so the rejection of Bohm's theory in favour of quantum 
mechanics---characterized via the Clifton-Bub-Halvorson (CBH) 
theorem \cite{CBH}
 in terms of 
three information-theoretic principles---involves taking the notion 
of quantum information as a new physical primitive. (By `information' 
here, I mean information in the physical sense, measured classically 
by the Shannon entropy and, in quantum mechanics, by the von Neumann 
entropy.) In `The CBH 
Characterization Theorem,' I outline the content of the CBH theorem. Finally, 
in `Quantum Information,' I argue that, 
just as Einstein's analysis (based on the assumption that we live in a 
world in which natural processes are subject to certain constraints 
specified by the principles of special relativity) 
shows that 
the mechanical structures in Lorentz's constructive theory 
(the ether, and the behaviour of electrons in the ether) 
are irrelevant to a physical explanation of electromagnetic phenomena, so the 
CBH analysis (based on the assumption that we live in a world in 
which there are certain constraints on the acquisition, 
representation, and communication of information) shows that the 
mechanical structures in Bohm's constructive theory (the guiding 
field, the behaviour of particles in the guiding field) are 
irrelevant to a physical explanation of
 quantum phenomena. You can, if you like, tell a 
story along Bohmian, or similar, lines (as in other `no collapse' 
interpretations) but, given the information-theoretic constraints, 
such a story can, in principle, have no excess empirical 
content over quantum mechanics (just as Lorentz's 
theory, insofar as it is constrained by the requirement to reproduce 
the empirical content of the principles of special relativity, 
can, in principle, have no excess empirical content over Einstein's theory).

\section{Principle vs Constructive Theories}

Einstein introduced the 
distinction between principle and constructive theories in an article
 on the significance of the special and 
general theories of relativity that he wrote for the London 
\textit{Times}, which appeared in the issue of November 28, 1919 
\cite{EinsteinTimes}:

\begin{quote}
    We can distinguish various kinds of theories in physics. Most of 
    them are constructive. They attempt to build up a picture of the 
    more complex phenomena out of the material of a relatively simple 
    formal scheme from which they start out. Thus the kinetic theory 
    of gases seeks to reduce mechanical, thermal, and diffusional 
    processes to movements of molecules---i.e., to build them up out 
    of the hypothesis of molecular motion. When we say that we have 
    succeeded in understanding a group of natural processes, we 
    invariably mean that a constructive theory has been found which 
    covers the processes in question. 
    
    Along with this most important class of theories there exists a 
    second, which I will call `principle theories.' These employ the 
    analytic, not the synthetic, method. The elements which form their 
    basis and starting-point are not hypothetically constructed but 
    empirically discovered ones, general characteristics of natural 
    processes, principles that give rise to mathematically formulated 
    criteria which the separate processes or the theoretical 
    representations of them have to satisfy. Thus the science of 
    thermodynamics seeks by analytical means to deduce necessary 
    conditions, which separate events have to satisfy, from the 
    universally experienced fact that perpetual motion is impossible.
\end{quote} 

Einstein's point was that 
relativity theory is to be understood 
as a principle theory. He returns to this theme in his 
`Autobiographical Notes' \cite[pp. 51--52]{EinsteinBiog},
 where he remarks that he first tried to find a 
constructive theory that would account for the known properties of 
matter and radiation, but eventually became convinced that the 
solution to the problem was to be found in a principle theory that
 reconciled the constancy of the velocity 
of light in vacuo for all inertial frames of reference, and the 
equivalence of inertial frames for all physical laws (mechanical 
as well as electromagnetic):

\begin{quote}
    Reflections of this type made it clear to me as long ago as 
shortly 
    after 1900, i.e., shortly after Planck's trailblazing work, that 
    neither mechanics nor electrodynamics could (except in limiting 
    cases) claim exact validity. By and by I despaired of the 
    possibility of discovering the true laws by means of constructive 
    efforts based on known facts. The longer and the more 
despairingly 
    I tried, the more I came to the conviction that only the 
    discovery of a universal formal principle could lead us to 
    assured results. The example I saw before me was thermodynamics. 
    The general principle was there given in the theorem: the laws of 
    nature are such that it is impossible to construct a 
    \textit{perpetuum mobile} (of the first and second kind). How, 
    then, could such a universal principle be found?
\end{quote}

A little later \cite[p. 57]{EinsteinBiog}, he adds:

\begin{quote}
    The universal principle of the special theory of relativity is 
    contained in the postulate: The laws of physics are invariant 
    with respect to the Lorentz-transformations (for the transition 
    from one inertial system to any other arbitrarily chosen system 
    of inertia). This is a restricting principle for natural laws, 
    comparable to the restricting principle for the non-existence of 
    the \textit{perpetuum mobile} which underlies thermodyamics.
\end{quote} 

According to Einstein, two very different sorts of 
theories should be distinguished in physics. One sort involves the 
reduction of a domain of 
relatively complex phenomena to the properties of simpler 
elements, as in the kinetic theory, which reduces
the mechanical and thermal behavior of gases to the motion of 
molecules, the elementary building blocks of the constructive theory. 
The other sort of theory is formulated in terms of `no go' principles 
that 
impose constraints on physical processes or events, as in 
thermodynamics (`no perpetual motion machines'). 
For an illuminating account of the role 
played by this distinction in Einstein's work, see the discussion by 
Martin Klein in \cite{Klein}.

The special theory of relativity is a principle theory, formulated in 
terms of two 
principles: the equivalence of inertial frames for all physical laws 
(the laws of electromagnetic phenomena as well as the laws of 
mechanics), and the constancy of the velocity of light in vacuo for 
all inertial frames. These principles are irreconcilable in the 
geometry of Newtonian space-time, where inertial frames are 
related by Galilean transformations. The required revision yields
Minkowski geometry, where inertial 
frames are related by Lorentz transformations. Einstein characterizes
the special principle of relativity, that the laws of physics are 
invariant with respect to Lorentz transformations from one inertial 
system to another, as `a restricting principle for 
natural laws, comparable to the restricting principle for the 
non-existence of the \textit{perpetuum mobile} which underlies 
thermodynamics.' (In the case of the general theory of relativity, 
the group of allowable transformations includes all differentiable 
transformations of the space-time manifold onto itself.) 
By contrast, the Lorentz theory \cite{Lorentz}, 
which 
derives the Lorentz transformation from the electromagnetic 
properties of the ether, 
and assumptions about the transmission of molecular forces through 
the ether, 
is a constructive theory. 

Consider the transition:

\smallskip
\noindent Lorentz's constructive mechanical theory of 
the electrodynamics of moving bodies 

\noindent $\longrightarrow$ Einstein's 
principle theory of special relativity 

\noindent $\longrightarrow$ Minkowski's formulation of 
Einstein's theory in terms of a non-Euclidean space-time geometry

\smallskip
\noindent Einstein showed that you could obtain a unified treatment of 
mechanical and electromagnetic phenomena---particles, electrons, light---by 
extending the idea of Galilean relativity (in a suitably modified 
form, involving the Lorentz transformation between inertial frames) 
to both mechanical and 
electromagnetic phenomena. In Minkowski's formulation of the theory,  
the relativistic principles are 
instantiated in a specific non-Newtonian geometry of 
space-time. In this new 
framework, rigid bodies are excluded by the symmetry group (i.e., 
they would transmit signals faster than light) and, strictly speaking, 
particles (insofar as they are small rigid bodies) are excluded. 
Instead, the field becomes the basic physical entity, as a new 
physical primitive. In particular, since an electromagnetic wave is not 
reduced to 
the vibratory motion of a mechanical medium (as a sound 
wave is reducible to the notion of air molecules), the ether is no 
longer required as the medium for the physical 
instantiation of an electromagnetic 
field. 

Now compare the historical transition:

\smallskip
\noindent Lorentz's constructive theory 

\noindent $\longrightarrow$ Einstein's 
principle theory 

\noindent $\longrightarrow$ Minkowski space-time

\smallskip 
\noindent with the transition in a modified version of Cushing's 
counterfactual history:

\smallskip
\noindent Bohm's constructive theory 

\noindent $\longrightarrow X$ 

\noindent $\longrightarrow$ 
Hilbert space quantum mechanics

\smallskip 
\noindent That is (for comparison with the relativistic case), 
assume that Bohm's theory was actually developed 
before Hilbert space quantum mechanics as a solution to some of the 
experimental difficulties of classical mechanics at the turn of the 
20'th century, and that there was an additional development, something 
like an Einsteinian 
formulation of quantum mechanics as a principle theory (the `$X$').

Without the 
`X' step, the Copenhagen 
argument for the completeness of Hilbert space quantum mechanics (and 
the associated rejection of Bohm's theory) in the 
counterfactual world seems implausible, as Cushing suggests. 
Similarly, to consider another counterfactual history, we might 
suppose that (after Lorentz's theory) the special theory of relativity was 
first formulated geometrically by Minkowski rather than Einstein, as 
an algorithm for relativistic kinematics and the Lorentz 
transformation, which is incompatible with the kinematics of Newtonian 
space-time. Without Einstein's analysis of the theory as a principle 
theory along the lines sketched above, 
it seems implausible to suppose that Lorentz's theory would 
have been dislodged by what would surely have seemed to be merely a 
convenient (but `counterintuitive and mind-boggling') algorithm.

In the following section, I argue that the missing `$X$' (in a 
logical sense) is supplied 
by the CBH characterization theorem for quantum 
theory in terms of three information-theoretic constraints and that, 
given this theorem, the relation between quantum mechanics 
and constructive theories like Bohm's theory should be seen as 
analogous to the relation between special relativity and Lorentz's 
theory. Just as special relativity involves a theory of the structure 
of space-time in which a field is a new physical primitive not 
reducible to the motion of a mechanical medium (ultimately, to the 
motion of particles), so quantum mechanics involves a 
theory of the algebraic structure of states and observables in which 
information is a new physical primitive not reducible to the behaviour 
of mechanical systems (the motion of particles or fields).

It should 
go without saying that I am not comparing the CBH theorem with 
Einstein's achievement in developing the special theory of relativity. 
To avoid any such suggestion, which would be ludicrous, let me say 
what would perhaps be a comparable achievement. Suppose, in a modified version 
of Cushing's counterfactual history, that in 1927 Bohm's theory was 
the dominant research paradigm in quantum physics. Suppose (in 1927) 
that CBH 
showed that one could dispense with the whole idea of a source-less field 
in configuration space guiding the motion of particles
by deriving the current Hilbert space theory from three 
information-theoretic constraints, and in terms of this (then new) Hilbert 
space theory also showed in detail how one could treat various quantum 
systems, formerly treated in terms of Bohm's theory, in a much 
simpler way, and in particular brought out the implications of 
entanglement as a new physical resource that could be exploited to 
develop novel 
forms of computation and cryptographic procedures that were 
impossible classically (cf. $E = mc^{2})$. 
In our actual history, since Hilbert space 
quantum mechanics and 
quantum information theory are already on the table, the CBH theorem is 
hardly 
more than a footnote to current theory. The purpose in pointing to the 
analogy is to argue that the relevance of the CBH 
theorem to the interpretative debate about Hilbert space quantum 
mechanics and the significance of constructive mechanical theories like 
Bohm's theory, 
is to be understood as 
similar to the relevance of Einstein's 
analysis of special relativity as a principle theory to Minkowski's
geometric formulation of the theory and Lorentz's constructive mechanical 
ether theory.

\section{The CBH Characterization Theorem}

The CBH characterization theorem is formulated in the general 
framework of $C^{*}$-algebras, which allows
 a mathematically
 abstract characterization of a physical theory that includes, as
 special cases, all classical mechanical theories of both wave and
 particle varieties, and all variations on quantum theory, including
 quantum field theories (plus any hybrids of these theories, such as 
 theories with superselection rules). So the analysis is not restricted to the
 standard quantum mechanics of a system represented on a single
 Hilbert space with a unitary dynamics, but is general 
 enough to cover
 cases of systems with an infinite number of degrees of freedom
  that arise in quantum field theory and the thermodynamic limit
 of quantum statistical mechanics (in which the number of
 microsystems and the volume they occupy goes to infinity, while the
 density defined by their ratio remains constant), including 
 the quantum theoretical
 description of exotic phenomena
 such as Hawking radiation, black hole evaporation, Hawking
 information loss, etc. The Stone-von Neumann theorem, 
 which guarantees the existence of a
 unique representation (up to unitary equivalence) of the canonical
 commutation relations for systems with a finite number of degrees of
 freedom, breaks down for such cases, and there will
 be many unitarily inequivalent
 representations of the canonical commutation relations. 
 One could, of course, consider weaker 
mathematical structures, but it seems that the $C^{*}$-algebraic 
machinery suffices for all physical theories that have been found to 
be empirically successful to date, including phase space theories and 
Hilbert space theories \cite{Landsmann}, and theories based on a 
manifold \cite{Connes}.

A $C^{*}$-algebra 
is essentially an abstract generalization of the structure of the 
algebra of operators on a Hilbert space. Technically, a (unital) 
$C^{*}$-algebra is a Banach $^{*}$-algebra over the 
complex numbers containing the identity, where the involution 
operation $^{*}$ and the norm are related by $\|A^{*}A\| = \|A\|^{2}$. 
So the algebra $\alg{B}(\hil{H})$ of all bounded operators on a 
Hilbert space 
$\hil{H}$ is a $C^{*}$-algebra, 
with $^{*}$  
the adjoint operation and $\|\cdot\|$ the 
standard operator norm. 

In standard quantum theory, a state on $\alg{B}(\hil{H})$ is
defined by
a density operator $D$ on $\hil{H}$ in terms of an expectation-valued
functional
$\rho(A)=\mbox{Tr}(AD)$ for all observables represented by
self-adjoint operators $A$ in $\alg{B}(\hil{H})$. This definition of $\rho(A)$
in terms of $D$
yields a positive normalized linear functional. So a
  state on a $C^{*}$-algebra $\alg{C}$ is defined, quite generally, as
any positive normalized
linear functional $\rho:\alg{C}\rightarrow\mathbb{C}$ on the algebra.
Pure states are defined by the condition that if
$\rho=\lambda\rho_{1}+(1-\lambda)\rho_{2}$ with $\lambda\in (0,1)$,
then
$\rho=\rho_{1}=\rho_{2}$;  other states are mixed.

The most general dynamical evolution of a system
represented by a $C^{*}$-algebra of observables is given by a
completely positive linear map $T$ on the algebra of observables,
where $0\leq T(I)\leq I$. The map or operation $T$ is called selective
if $T(I)<I$
and nonselective if $T(I)=I$. A yes-no measurement of some idempotent
observable represented by a projection operator $P$ is an example of a
selective operation. Here,  $T(A)=PAP$
for all $A$ in the $C^{*}$-algebra $\alg{C}$,
and $\rho^{T}$, the transformed (`collapsed') state, is the final state
obtained after measuring $P$ in the
state $\rho$
  and ignoring all elements of the ensemble that do not
yield the eigenvalue 1 of $P$
(so $\rho^{T}(A)=\rho(T(A))/\rho(T(I))$ when $\rho(T(I))\not=0$, and
$\rho^{T}=0$ otherwise). The time evolution in the Heisenberg picture
induced by a unitary operator $U\in\alg{C}$ is an example of a
nonselective operation. Here, $T(A)=UAU^{-1}$. Similarly, the
measurement of an observable $O$ with spectral measure $\{P_{i}\}$,
without selecting a particular outcome, is an example of a
nonselective operation, with $T(A) = \sum _{i=1}^{n}P_{i}AP_{i}$. Note 
that any completely positive linear map can be regarded as the 
restriction to a local 
system of a unitary map on a larger system.

A 
representation of a $C^{*}$-algebra $\alg{C}$ is any mapping 
$\pi:\alg{C}\rightarrow\alg{B}(\hil{H})$ that preserves the linear, 
product, and $^{*}$ structure of \alg{C}. The representation is faithful
if $\pi$ is 
one-to-one, in which case $\pi(\alg{C})$ is an isomorphic copy of 
$\alg{C}$. The Gelfand-Naimark theorem says that every abstract $C^{*}$-algebra 
has a concrete 
faithful representation as a norm-closed $^{*}$-subalgebra of 
$\alg{B}(\hil{H})$, for some appropriate Hilbert space $\hil{H}$. 
In the case of systems with an 
infinite number of degrees of freedom  (as in quantum field theory), 
there are inequivalent representations of the 
$C^{*}$-algebra of observables defined by the commutation relations.

The relation between classical theories and $C^{*}$-algebras is 
this: every 
\textit{commutative} $C^{*}$-algebra $\alg{C}$ is isomorphic to the set 
$C(X)$ of 
all continuous complex-valued functions on a locally 
compact Hausdorff space 
$X$. If $\alg{C}$ has a multiplicative identity, $X$ is compact. 
So behind every abstract abelian $C^{*}$-algebra there is a 
classical phase space theory defined by this `function representation' 
on the phase space $X$. Conversely, every classical phase space theory 
defines a $C^{*}$-algebra. For example, the observables of a classical 
system of $n$ particles---real-valued functions on the phase space 
$\mathbb{R}^{6n}$---can be represented as the self-adjoint elements of 
the $C^{*}$-algebra $\alg{B}(\mathbb{R}^{6n})$ of all continuous complex-valued 
functions $f$ on $\mathbb{R}^{6n}$. The phase space 
$\mathbb{R}^{6n}$ is locally compact and can be made compact by 
adding just one point `at infinity,' or we can simply consider a 
closed and bounded (and thus compact) subset of $\mathbb{R}^{6n}$. 
The statistical states of 
the system are given by probability 
measures $\mu$ on 
$\mathbb{R}^{6n}$, and pure states, corresponding to maximally complete 
information about the particles, are given by
the individual points of 
$\mathbb{R}^{6n}$. The system's state $\rho$ in 
the $C^{*}$-algebraic sense is the 
expectation functional corresponding to $\mu$, defined by
$\rho(f)=\int_{\mathbb{R}^{6n}}f\mbox{d}\mu$. 

So classical theories are characterized  by commutative
$C^{*}$-algebras. CBH identify quantum theories with a certain 
subclass of noncommutative $C^{*}$-algebras; specifically, theories 
where 
(i) the observables of the theory are represented by the self-adjoint 
 operators in a 
 noncommutative $C^{*}$-algebra (but the algebras of observables of 
 distinct systems commute), (ii) the states of the theory are represented 
 by $C^{*}$-algebraic states (positive normalized linear functionals 
 on the $C^{*}$-algebra), and spacelike separated systems can be 
 prepared in
entangled states that allow what Schr\"{o}dinger \cite[p. 556]{Schr1} 
calls `remote 
steering', and (iii) dynamical 
changes are 
represented by 
 completely positive linear maps. For example, the standard quantum 
 mechanics of a system with a finite number of degrees of freedom 
 represented on a single Hilbert space with a unitary dynamics 
 defined by a given Hamiltonian is a 
 quantum theory, and theories with different Hamiltonians can be considered 
 to be empirically inequivalent quantum theories. Quantum field theories 
 for systems with an infinite number 
 of degrees of freedom, where there are many unitarily 
 inequivalent Hilbert space representations of the canonical 
 commutation relations, are quantum theories.  
 (For a detailed discussion and motivation for 
this identification, see \cite{BubWhy, CBH, Halvorson2, HalvorsonBub}.)

What CBH showed was that one can derive the basic 
kinematic features of a quantum-theoretic 
 description of physical systems in the above sense 
 from three fundamental information-theoretic 
 constraints: (i) the impossibility of superluminal information transfer 
 between two physical systems by performing measurements on one of 
 them, (ii) the impossibility of perfectly broadcasting the information 
 contained in an unknown physical state (for pure states, this amounts 
 to `no cloning'), and (iii) the impossibility of communicating information 
so as to implement a certain 
 primitive cryptographic protocol, called `bit commitment,'
  with unconditional security. They also partly demonstrated the 
  converse derivation, leaving open a question concerning nonlocality 
  and  bit commitment. This remaining
 issue has been resolved by Hans Halvorson \cite{Halvorson1}, so we have a 
  characterization theorem for quantum theory in terms of the three 
  information-theoretic constraints. 
  
To clarify the significance of the information-theoretic 
  constraints, consider a 
composite quantum system A+B, consisting of two subsystems, A and B. 
For simplicity, assume the systems are identical, so their 
$C^{*}$-algebras $\alg{A}$ and $\alg{B}$ are isomorphic. 
The 
observables of the component systems A and B are 
represented by the self-adjoint elements of $\alg{A}$ and $\alg{B}$, 
respectively. Let  $\alg{A}\vee\alg{B}$ denote the $C^{*}$-algebra 
generated by $\alg{A}$ and $\alg{B}$. The physical states of A, B, and 
A+B, are given by positive normalized linear functionals on their 
respective algebras that encode the expectation values of all 
observables (cf. standard quantum theory, where a state on 
$\alg{B}(\hil{H})$ is 
defined by
a density operator $D$ on $\hil{H}$ in terms of an expectation-valued 
functional  
$\rho(A)=\mbox{Tr}(AD)$ for all observables represented by 
self-adjoint operators $A$ in $\alg{B}(\hil{H})$.) 
To capture the idea that A and B are \textit{physically 
distinct} systems, 
CBH make the assumption that any state of $\alg{A}$ is
compatible with any state of $\alg{B}$, i.e., for any state
$\rho _{A}$ of $\alg{A}$ and $\rho _{B}$ of $\alg{B}$,
there is a state $\rho$ of $\alg{A}\vee \alg{B}$ such that $\rho
|_{\alg{A}}=\rho _{A}$ and $\rho |_{\alg{B}}=\rho _{B}$.

The sense of the `no superluminal information transfer via measurement' 
constraint 
is 
that when Alice and Bob, say, perform local measurements, Alice's 
measurements can have no influence on the statistics for the outcomes 
of Bob's measurements, and conversely. That is, merely performing
 a local measurement cannot, in and of 
itself, convey any information to a physically distinct system, so 
that everything `looks the same' to that system after the 
measurement operation as before, in terms of the expectation values 
for the outcomes of measurements. CBH show that it 
follows from this constraint that 
A and B are 
\textit{kinematically independent} systems if they are physically 
distinct in the above sense, i.e., every element of $\alg{A}$ 
commutes pairwise with every element of $\alg{B}$.

The `no broadcasting' condition now ensures that the individual algebras  
$\alg{A}$ and  $\alg{B}$ are noncommutative. Broadcasting is a 
process closely related to cloning. In fact, for pure states, broadcasting 
reduces to cloning.  In cloning, a ready state $\sigma$ of 
a system B and the state to be cloned $\rho$ of system A are 
transformed into two copies of $\rho$. In broadcasting, a ready state 
$\sigma$ of B and the state to be broadcast $\rho$ of A are transformed to a new 
state $\omega$ of A+B, where the marginal states of $\omega$ with 
respect to both A and B are $\rho$. In elementary quantum 
mechanics, neither cloning nor broadcasting is possible in general. A 
pair of pure states can be cloned if and only if they are orthogonal 
and, more generally, a pair of mixed states can be broadcast if 
and only if they are represented by mutually commuting density 
operators. CBH show that broadcasting and cloning are always possible for 
classical systems, i.e., in the commutative case there is a 
universal broadcasting map that clones 
any pair of input pure states and broadcasts any pair of input mixed 
states. Conversely, they show that if any two states can be (perfectly) 
broadcast, 
then any two 
pure states can be cloned; and if two pure states of a $C^{*}$-algebra 
can be cloned, then they must be orthogonal. So, if any 
two states can be broadcast, then all pure states are orthogonal, 
from which it follows that the algebra is commutative.

The quantum mechanical phenomenon of interference  is the 
physical manifestation of the noncommutativity of quantum observables 
or, equivalently, the superposition 
of quantum states. 
So the impossibility of perfectly broadcasting the 
information contained in an unknown physical state, or of cloning or 
copying the information in an unknown pure state, is the 
information-theoretic counterpart of interference. 

Now, if $\alg{A}$ and $\alg{B}$ are noncommutative and mutually commuting 
algebras of observables associated with distinct spatially separated 
systems, 
it can be shown that there are nonlocal entangled states on the
$C^{*}$-algebra
$\alg{A}\vee\alg{B}$ they generate (see \cite{Landau, Summers, 
Bacciagaluppi}, and---more relevantly here, in 
terms of a specification of the range of entangled states that can be 
guaranteed to exist---\cite{Halvorson1}). So it seems that 
entanglement---what Schr\"{o}dinger \cite[p. 555]{Schr1} called
 `\textit{the} characteristic trait of
quantum mechanics, the one that enforces its entire departure from
classical lines of thought'--- follows automatically in 
any theory with a noncommutative algebra of observables. That is, it 
seems that once we assume `no superluminal information transfer via 
measurement,' and `no broadcasting,' the class of allowable physical 
theories is restricted to those theories in which physical systems manifest 
both interference \textit{and} nonlocal entanglement. But this 
conclusion is surely too quick, since the derivation of entangled 
states depends on formal
properties of the
$C^{*}$-algebraic machinery. Moreover, we have no assurance that two 
systems in an entangled state will
maintain their entanglement indefinitely as 
they separate in space, which is the case for quantum entanglement. 
But this is precisely what is required by
the cheating 
strategy that thwarts
secure bit commitment, since Alice will have to keep one system of 
such a pair and send the other system to Bob, whose degree of spatial separation 
from Alice is irrelevant, in principle, to the implementation of the protocol. 
In an information-theoretic characterization of quantum theory, the 
fact that entangled states of composite systems can be instantiated, 
and instantiated nonlocally so that the entanglement of composite systems 
is maintained as the subsystems separate in space, should be shown to 
follow from some information-theoretic principle. The role of the `no bit
commitment' constraint is to guarantee entanglement maintenance over 
distance, that is, the existence of a certain class of nonlocal 
entangled states---hence it gives us nonlocality, not merely 
`holism.' 

Bit commitment is a cryptographic protocol in which one party, Alice,
supplies an encoded bit
to a second party, Bob, as a warrant for her commitment to 0 or 1. 
The information available in the encoding
should be insufficient for Bob to ascertain the value of the bit at 
the initial commitment stage, but
sufficient, together with further information supplied by Alice at a
later stage when she is supposed to `open' the commitment by revealing the
value of the bit, for Bob to be convinced that the protocol does not
allow Alice to cheat by encoding the bit in a way that leaves her
free
to reveal either 0 or 1 at will.

In 1984, Bennett and Brassard \cite{BB} proposed a quantum bit
commitment protocol now referred to as BB84. The basic idea was to
encode
the 0 and 1
commitments as two quantum mechanical mixtures
represented by the same density operator, $\omega$. As they showed,
Alice can cheat by adopting an 
EPR
attack or cheating strategy. Instead of following the protocol and 
sending a particular mixture to Bob she prepares pairs of
particles A+B in the same entangled state $\rho$, where 
$\rho |_{\alg{B}} = \omega$. She keeps one of each pair (the ancilla A) 
and sends the second
particle B to Bob, so that Bob's particles are in the mixed state 
$\omega$. In this way she can reveal either bit at will at the opening stage, by
effectively steering Bob's particles into
the desired mixture via appropriate measurements on her ancillas. Bob
cannot detect this cheating strategy.

Mayers \cite{Mayers1, Mayers2}, and Lo and Chau \cite{Lo}, showed that
the insight of
Bennett and Brassard
can be extended to a proof that a generalized version of
the EPR cheating strategy can always be applied, if the
Hilbert space is enlarged in a suitable way by introducing additional
ancilla particles. The proof of this `no go' quantum bit commitment
theorem
exploits biorthogonal decomposition
via a result by Hughston, Jozsa, and Wootters \cite{HJW}. 
Informally, this  says that for a quantum
mechanical system
consisting of two (separated) subsystems represented by
 the $C^{*}$-algebra
$\alg{B}(\hil{H}_{1}) \otimes \alg{B}(\hil{H}_{2})$, any mixture of
states on $\alg{B}(\hil{H}_{2})$ can be generated
from a distance by performing an appropriate POV-measurement on the
system represented by $\alg{B}(\hil{H}_{1})$, for an appropriate
entangled state of the composite system
 $\alg{B}(\hil{H}_{1}) \otimes \alg{B}(\hil{H}_{2})$. 
Schr\"{o}dinger \cite[p. 556]{Schr1} 
 called this `remote steering' and found the possibility so physically 
 counterintuitive that he speculated \cite[p. 451]{Schr2} 
 (wrongly, as it turned out) 
 that experimental 
 evidence would eventually show that this was simply an artifact 
 of the theory, and that any entanglement between two systems 
 would spontaneously (and instantaneously) decay 
 as the systems separated in space.
Remote steering is what makes it possible for Alice to cheat in her bit
commitment protocol with Bob. It is easy enough to see this for the
original BB84 protocol. Suprisingly, this is also the case for
 any conceivable quantum bit commitment
protocol. (See Bub \cite{BubBC} for a discussion.)

Now, unconditionally secure bit commitment is also impossible for
classical
systems, in which the algebras of observables are 
commutative.\footnote{Adrian Kent \cite{Kent} has shown 
how to implement a secure classical 
bit commitment protocol by exploiting relativistic signalling 
constraints in a timed sequence of communications between verifiably 
separated sites for both Alice and Bob. In a bit commitment protocol, 
as usually construed, there is a time interval 
of arbitrary 
length, where no information is exchanged, 
between the end of the commitment stage of the protocol and 
the opening or unveiling stage, when Alice reveals the value of the 
bit. 
 Kent's ingenious scheme effectively involves a third stage between the 
commitment state and the unveiling stage, in which information is 
exchanged between Bob's sites and Alice's sites at regular intervals 
until one of Alice's sites 
chooses to unveil the originally committed bit. At this moment of 
unveiling the protocol is not yet complete, because a further sequence of 
unveilings is required between Alice's sites and corresponding sites 
of Bob before Bob has all the information required to verify the 
commitment at a single site. If a bit commitment protocol 
is understood to 
require an arbitrary amount of `free' time between the end of the 
commitment stage and the opening stage (in which no step is to be 
executed in the protocol), then unconditionally secure bit commitment 
is impossible for classical systems.  
(I am indebted to Dominic Mayers for clarifying this 
point.)} But the insecurity of any bit commitment protocol
in a noncommutative setting depends on considerations entirely
different from those in a classical commutative setting. Classically,
unconditionally secure bit commitment is impossible, essentially
because
 Alice can send (encrypted) information to Bob that guarantees the
 truth of an exclusive classical disjunction (equivalent
to her commitment to a 0 or a 1) only if the information is biased
towards one
of the alternative disjuncts (because a classical exclusive
disjunction is true
if and only if one of the disjuncts is true and the other false). No
principle of classical mechanics precludes Bob from extracting this
information.
So the security of the protocol cannot be unconditional and
 can only depend on issues of
computational complexity.

 By contrast, the noncommutativity of quantum mechanics allows the 
 possibility of different mixtures associated with
the same density operator. 
 If Alice sends Bob one of two mixtures associated with the same
 density operator to establish her commitment, then she
is, in effect, sending Bob evidence for the
truth of an exclusive disjunction that is not based on the selection
of a particular disjunct (`0 or 1').
What thwarts the possibility of using the
ambiguity of mixtures in this way to implement an unconditionally
secure bit commitment protocol is the existence of nonlocal entangled
states, and the maintenance of entanglement as entangled systems 
separate. This allows Alice to cheat by preparing
a suitable entangled state instead of one of the mixtures, where the
reduced density operator for Bob is the same as that of the mixture.
Alice is
then able to steer Bob's systems remotely into either of the two mixtures
associated with the alternative commitments at will. (See Bub \cite{BubWhy} 
for a further discussion.)

So what \textit{would} allow unconditionally secure bit commitment in a
noncommutative theory is the spontaneous decay of entangled states of 
composite systems as the component systems separate in space 
(specifically, the entangled state of the pair of systems prepared by 
Alice, one of which she sends to Bob).
One can therefore take Schr\"{o}dinger's remarks (with hingdsight) 
as relevant to the
question of whether or not secure bit
commitment is possible in our world. In effect, Schr\"{o}dinger
raised the possibility
that we live in a quantum-like world in which  secure bit commitment
is possible! It follows that the 
impossibility of unconditionally secure bit commitment entails that, 
for any mixed state that Alice and Bob can prepare by following some 
(bit commitment) protocol, there is a corresponding
entangled state that remains entangled as the component systems 
separate and pass between Alice and Bob. 

To sum up: the content of the CBH theorem is that a quantum theory---a theory
 where
(i) the observables of the theory are represented by the self-adjoint 
 operators in a 
 noncommutative $C^{*}$-algebra (but the algebras of observables of 
 distinct systems commute), (ii) the states of the theory are represented 
 by $C^{*}$-algebraic states (positive normalized linear functionals 
 on the $C^{*}$-algebra), and spacelike separated systems can be 
 prepared in
entangled states that allow remote steering, and (iii) dynamical 
changes are 
represented by 
 completely positive linear maps---can be characterized by the 
three information-theoretic `no-go's': no superluminal communication 
of information via measurement, no (perfect) broadcasting, and no 
(unconditionally secure) bit commitment. 

\section{Quantum Information}

The significance of the CBH theorem is that we can now see quantum 
mechanics as a principle theory, where the principles are 
information-theoretic constraints. A relativistic theory is a theory 
characterized by certain 
symmetry or invariance properties, defined in terms of a group of 
space-time transformations. Following Einstein's formulation of 
special relativity as a principle theory, we understand this 
invariance to be a consequence of the fact that we live in a world in 
which natural processes are subject to certain constraints. (Recall 
Einstein's characterization of the special principle of relativity as 
`a restricting principle for natural laws, comparable to the 
restricting principle of the non-existence of the \textit{perpetuum 
mobile} which underlies thermodynamics.') CBH treat a quantum theory as a 
theory in which the observables and states have a certain 
characteristic algebraic structure. So for CBH, `quantum' is a 
structural adjective applicable to theories, just as `relativistic' is.
Unlike relativity theory, quantum 
mechanics was born as a recipe or algorithm for caclulating the 
expectation values of observables measured by macroscopic measuring 
instruments. The interpretative problems arise because
 this Hilbert space theory has no phase 
space representation. Without Einstein's analysis, we could also 
see Minkowski space-time simply as an algorithm for relativistic kinematics 
and the Lorentz transformation, which is incompatible with the 
kinematics of Newtonian space-time. What Einstein's analysis provides is a 
rationale for taking the structure of space-time as Minkowskian: we 
see that this is required for the consistency of the two principles of 
special relativity. From 
this perspective, it is also clear that, insofar as a constructive theory like 
Lorentz's theory is constrained by the requirement to reproduce the 
empirical content of the principles of special relativity (which 
means that the ether as a rest frame for electromagnetic phenomena 
must, in principle, be undetectable), such a theory can have no excess 
empirical content over special relativity. Cushing \cite[p. 
193]{Cushing2} 
quotes Maxwell as asking whether `it is not more philosophical to 
admit the existence of a medium which we cannot at present perceive, 
than to assert that a body can act at a place where it is not.' Yes, 
but not if we also have to admit that, in principle, as a matter of 
physical law, if we live in a world in 
which events are constrained by the two relativistic principles, the medium 
must remain undetectable.

Consider again the transition:

\smallskip
\noindent Lorentz's constructive theory 

\noindent $\longrightarrow$ special relativity 
as a principle theory (via Einstein's analysis)

\noindent $\longrightarrow$ Minkowski space-time

\smallskip
\noindent and the counterfactual history:

\smallskip
\noindent Bohm's constructive theory 

\noindent $\longrightarrow$ quantum mechanics as a principle theory 
(via CBH) 

\noindent $\longrightarrow$ 
Hilbert space representation of states and observables

\smallskip

What the CBH analysis provides is a rationale for taking the 
structure of states and observables associated with quantum phenomena 
as a noncommutative $C^{*}$-algebra, 
represented on a Hilbert space with no phase space representation.
From the CBH theorem, a theory satisfies the information-theoretic 
constraints if and only if it is empirically equivalent to a quantum 
theory (a theory where the observables, the states, and the dynamics 
are represented as outlined at the end of Section 3). 
So if the information-theoretic 
constraints are satisfied, a constructive theory like Bohm's theory 
can have no excess empirical content over a quantum theory. 
Just as in the case of Lorentz's theory, Bohm's theory will have to 
posit contingent assumptions to hide the additional mechanical 
structures (the hidden variables will have to remain hidden), 
so that \textit{in principle}, as a matter of physical law, 
there \textit{could not be} any evidence favouring the 
theory over quantum theory. 

Consider how this is achieved in Bohm's theory. The additional 
mechanical 
structures in Bohm's theory are the particle trajectories in 
configuration space, and the wave function as a guiding field. 
The dynamical evolution of a 
Bohmian particle is described by a deterministic 
equation of motion in configuration space that is guaranteed to produce the 
quantum statistics for all quantum measurements, 
if the initial distribution over particle positions (hidden variables)
is the Born distribution. The Born distribution is treated as an 
equilibrium distribution, and non-equilibrium distributions can be 
shown to 
yield predictions that conflict with the information-theoretic 
constraints.
Valentini \cite{Valentini} shows how non-equilibrium distributions can be
associated with 
such phenomena as instantaneous signalling between spatially 
separated systems and the 
possibility of distinguishing nonorthogonal pure states 
(hence the possibility of cloning such states). Key distribution 
protocols whose security depends on `no information gain without 
disturbance' and `no cloning' would then be insecure against attacks 
based on exploiting such non-equilibrium distributions. 

On Bohm's theory, 
the explanation for the fact that the information-theoretic constraints 
hold in our universe 
is that the universe 
has in fact reached the equilibrium 
state with respect to the distribution of hidden variables. 
But now it is clear that there can be no empirical evidence for the 
additional structural elements of Bohm's theory that would represent excess 
empirical content over a quantum theory, because such evidence 
is in principle unobtainable in the equilibrium state. If the 
information-theoretic constraints apply at the phenomenal 
level then, according to Bohm's theory, the universe must be in 
the equilibrium state, and in that case there can be no phenomena 
that are not part of the empirical content of a quantum 
theory (i.e., the statistics of quantum superpositions and entangled states). 
Since a similar analysis will apply to 
any `no collapse' hidden variable theory---this, in effect, is what 
the `no go' hidden variable theorems tell us: any such theory will 
have to incorporate the basic features of 
Bohm's theory---the additional non-quantum structural elements that 
these theories postulate cannot be doing any work in providing a 
physical explanation of quantum phenomena that is not 
already provided by an empirically equivalent quantum theory.

Of course, it could be the case that we are mistaken about 
the information-theoretic 
constraints, and that some day we will find 
  experimental evidence that conflicts with the predictions of a quantum 
  theory. The above claim about constructive theories like Bohm's theory 
  is a conditional claim  
  about what follows if the information-theoretic constraints 
  \textit{do} in fact hold in our world. To put the point 
  differently: an acceptable mechanical theory of quantum phenomena 
  that includes an account of our measuring instruments as well as 
  the phenomena they reveal \textit{must violate at least one of the 
  information-theoretic constraints}. 
  
What led to Lorentz's theory was a problem about the 
electromagnetic field, conceived as an aspect of the motion of a 
mechanical medium. The rejection of Lorentz's constructive theory in 
favour of Einstein's principle theory  requires that we consider a 
field as a new physical primitive, not 
reducible to the motion of particles or a mechanical medium. 
What led to Bohm's theory was a problem about the 
difficulty of representing information from macroscopic classically 
described measuring 
instruments in a phase space theory that could account for the 
behaviour of the 
measuring instruments as well as the phenomena revealed by these 
instruments. If the $C^{*}$-algebra is commutative, there is a phase space 
representation of the theory---not necessarily the phase space of 
classical mechanics, but a theory in which the observables of the 
$C^{*}$-algebra are replaced by `beables' (Bell's term, see \cite{Bell}) 
or dynamical quantities, 
and the $C^{*}$-algebraic states are replaced by states representing complete 
catalogues of properties (idempotent quantities). In this case, it is 
possible to extend the theory to include the measuring instruments 
that are the source of the $C^{*}$-algebraic statistics, so that they are no 
longer `black boxes' but constructed out of systems that are 
characterized by properties and states of the phase space theory. 
That is, the $C^{*}$-algebraic theory can be replaced by a `detached observer' 
theory of the 
physical processes underlying the phenomena, to use Pauli's term 
\cite{Pauli}, 
including the processes involved in the functioning of measuring 
instruments. Note that this depends on a representation theorem. In the 
noncommutative case, we are guaranteed only the existence of a 
Hilbert space representation of the $C^{*}$-algebra, and the 
possibility of a `detached observer' description of the phenomena is 
a further question to be investigated. 

In a review of Cushing's \cite{Cushing2}, di Salle \cite[p. 755]{DiSalle} 
quotes Pauli as remarking in his \textit{Theory of Relativity} 
\cite[p. vi]{Pauli} that the ether `had to 
be given up, not only because it turned out to be unobservable, but 
because it became superfluous as an element of a mathematical 
formalism, the group-theoretical properties of which would only be 
disturbed by it.' Similarly, Pauli says, the concept of definite particle 
trajectories or space-time orbits had to be given up in quantum 
mechanics `not only because the orbits are unobservable, but 
because they became superfluous and woud disturb the symmetry inherent 
in the general transformation group underlying the mathematical 
formalism of the theory.' DiSalle comments:
\begin{quote}
	Evidently this is neither a simple empiricist rejection of the 
	unobservable, nor an operationalist reduction of the meanings of 
	theoretical terms to processes of measurement. Instead, it asserts 
	that the purpose of any formalism in physics is to represent the 
	known lawful behaviour of observable systems, and that distinctions 
	or symmetries that don't belong to observable systems don't belong to 
	their theoretical representation either. \ldots Thus it seems odd 
	that Cushing should ask, `What is it about the formalism of quantum 
	mechanics that makes it so difficult to tell a story that we feel we 
	understand about physical phenomena?' (p.341). The problem isn't 
	with the formalism at all; the project of telling such a story is 
	orthogonal to that of the formalism, which is to represent the 
	structure of the physical world as it actually reveals itself to us. 
	That structure may appear bizarre, but its bizarre aspects are 
	necessarily incorporated into deterministic alternatives to quantum 
	mechanics. Moreover, to regard the uncertainty relations as a kind 
	of natural `conspiracy' to hide the underlying determinism must have 
	seemed, to Pauli \textit{et al.}, precisely as odd as accepting the 
	Lorentz contraction instead of special relativity, and for precisely 
	the same reason. So, instead of Cushing's question, one could ask, 
	what is it about physical reality that makes it difficult to 
	represent it by a deterministic theory? Why is it that any 
	empirically adequate deterministic theory must be so constructed as 
	to mimic an indeterministic theory in every conceivable empirical 
	circumstance?
	\end{quote}

DiSalle's question is 
answered by the CBH characterization of quantum mechanics in terms of 
information-theoretic principles. 
The rejection of `detached observer' hidden variable theories
in favour of quantum 
mechanics requires that our 
measuring instruments \textit{ultimately remain black 
boxes} at some level. That is, a quantum description will
have to introduce a `cut' between what we take to be the ultimate measuring 
instrument in a given measurement process
and the quantum phenomenon revealed by the instrument, which means that 
the ultimate measuring instrument is treated simply as a probabilistic 
source of a range of labelled events or `outcomes,' i.e., effectively 
as a source of information in Shannon's sense. But this amounts 
to treating quantum mechanics as \textit{a theory about the 
  representation and manipulation of information} constrained by 
  the possibilities and 
impossibilities of information-transfer in our world (a fundamental 
  change in the aim of physics), rather than a theory 
  about the behavior of nonclassical waves and particles. 
  
Something like this view 
seems to be implicit in Bohr's complementarity interpretation of 
quantum theory. For 
Bohr, quantum mechanics is complete and there is no measurement 
problem, but measuring instruments ultimately remain outside the 
quantum description: the placement of the `cut' between system and measuring 
instrument is arbitrary, but the cut must be placed somewhere. 
Similarly, the argument here is that, if the information-theoretic 
constraints hold in our world, the measurement problem is a 
pseudo-problem, and the whole idea of an empirically equivalent 
`interpretation' of quantum 
theory that `solves the measurement problem' is to miss the point of 
the quantum revolution.
  
So a consequence of rejecting 
`detached observer' hidden variable
theories is that we recognize information as a new physical 
primitive, not reducible to the properties of particles or fields. 
An entangled 
state should be thought of as a new sort of nonclassical communication channel 
that we have discovered to exist in our universe, i.e., as a new sort of `wire.' 
We can use 
these communication channels to do things that would be impossible 
otherwise, e.g., to teleport states, to compute in new ways, etc. Quantum 
theory is then about the properties of these communication channels, and 
about the representation and manipulation of states as sources of information
in this physical sense.

The question: `What is information in the physical sense (if it is not 
reducible to the properties of particles or fields)?' should be seen as 
like the question: 
`What is a field in the physical sense (if it is not reducible to the 
motion of particles or a mechanical medium)?' The answer is something like this: 
Quantum mechanics 
represents the discovery that there are new sorts of information 
sources and communication 
channels in nature (represented by quantum states), and the theory is about the 
properties of these information sources and communication channels. 
You can, if you like, 
tell a mechanical story about quantum phenomena (via 
Bohm's theory, for example) but such a story, if constrained by the 
information-theoretic principles, can have no excess 
empirical content over quantum mechanics, and the additional 
non-quantum structural elements will be explanatorily superfluous. 
So the mechanical story for quantum phenomena is like an ether story 
for electromagnetic fields. Just as 
the ether story attempts to make sense of the behaviour of fields by 
proposing an ether that is a sort of sui generis mechanical system 
different from all other mechanical systems, so a Bohmian story 
attempts to make sense of quantum phenomena by introducing a field 
(the quantum potential or guiding field) that is a sort of sui 
generis field different from other physical fields.\footnote{In fact, 
in \cite[section 3.2]{BohmHiley} Bohm and Hiley 
suggest that the guiding field should be understood as a sort of 
informational field.}

Cushing \cite[p. 204]{Cushing2} quotes Lorentz 
(from the conclusion of the 1916 edition of 
\textit{The Theory of Electrons}) as complaining that `Einstein simply 
postulates what we have deduced.'

\begin{quote}
I cannot speak here of the many highly interesting applications which 
Einstein has made of this principle [of relativity]. 
His results concerning electromagnetic and optical phenomena ... agree in 
the main with those which we have obtained in the preceding pages, 
the chief difference being that Einstein simply postulates what we have deduced, 
with some difficulty and not altogether satisfactorily, 
from the fundamental equations of the electromagnetic field. By doing 
so, he may certainly take credit for making us see in the negative 
result of experiments like those of Michelson, Rayleigh and Brace, 
not a fortuitous compensation of opposing effects, but the 
manifestation of a general and fundamental principle. 

Yet, I think, something may also be claimed in favour of the form in which 
I have presented the theory. I cannot but regard the aether, which can be 
the seat of an electromagnetic field with its energy and its vibrations, 
as endowed with a certain degree of substantiality, however different 
it may be from all ordinary matter. In this line of thought, it seems natural 
not to assume at starting that it can never make any difference whether a 
body moves through the aether or not, and to measure distances and lengths 
of time by means of rods and clocks having a fixed position relative to 
the aether.
\end{quote}

Similarly, one might complain that CBH simply postulate what is 
ultimately \textit{explained} by a Bohmian (or other `no collapse') theory. 
Just as the rejection of Lorentz's complaint involves taking the 
field as a new physical primitive (tantamount to `no ether,' which 
follows once we accept the principles of special relativity 
as basic to an explanatory account of electromagnetic phenomena), 
so the rejection of the analogous 
complaint in the quantum case involves taking  
information as a new physical primitive (tantamount to `measuring 
instruments are ultimately to be treated as black boxes,' 
which follows---via the `no go' theorems---once we 
accept the three information-theoretic constraints as basic to an 
explanatory account of quantum phenomena). 

To conclude, it might be worthwhile clarifying what is \textit{not} 
being argued here. Firstly, the CBH theorem should not be understood 
as providing a `constructive' explanation for the quantum formalism, 
along the lines suggested by Chris Fuchs \cite{Fuchs} (or the 
axiomatization proposed by 
Lucien Hardy \cite{Hardy}, or by quantum logicians), but rather as a `principled' 
reconstruction of the theory within a suitably general mathematical 
framework. Secondly, the claim that quantum mechanics is about 
quantum information---that quantum mechanics is a \textit{principle 
theory} of information (in the sense in which Einstein regarded 
special relativity as a principle theory)---and that this physical 
notion of information is not reducible to the properties of particles 
or fields, is not to be construed as the claim that quantum mechanics 
is about observers and their epistemological concerns, nor that 
we have derived `it from bit' in Wheeler's sense \cite{Wheeler}
of a `participatory 
universe,' nor that the basic stuff of the world is informational in 
an intentional sense. (Recall Shannon's remark \cite[p.31]{Shannon}
that `the semantic 
aspects of communication are irrelevant to the engineering 
problem'---it is the `engineering' sense of information that is 
relevant to the CBH theorem.) Rather, the 
claim is that the lesson of modern physics is that a 
principle theory is the 
best one can hope to achieve as an explanatory account of quantum 
phenomena.

\section*{Acknowledgements}

The ideas in this paper originated during a research leave supported 
 by a University of Maryland Sabbatical Leave Fellowship and 
 a General Research Board Fellowship in 2001--2002.

\end{document}